# Comparative Evaluation of Statistical Orbit Determination Algorithms for Short-Term Prediction of Geostationary and Geosynchronous Satellite Orbits in NavIC Constellation

T. V. Ramanathan[* (1)] and R. A. Chipade[(1, 2)]

[1]Department of Statistics, Savitribai Phule Pune University, Pune – 411007; e-mail: ram@unipune.ac.in

[2]Space Applications Centre (ISRO), Ahmedabad – 380015; e-mail: radhikachipade@gmail.com

## Abstract

NavIC is a newly established Indian regional Navigation Constellation with 3 satellites in geostationary Earth orbit (GEO) and 4 satellites in geosynchronous orbit (GSO). Satellite positions are essential in navigation for various positioning applications. In this paper, we propose a Bootstrap Particle Filter (BPF) approach to determine the satellite positions in NavIC constellation for short duration of 1 hr. The Bootstrap Particle filter-based approach was found to be efficient with meter level prediction accuracy as compared to other methods such as Least Squares (LS), Extended Kalman Filter (EKF), Unscented Kalman Filter (UKF) and Ensemble Kalman Filter (EnKF). The residual analysis revealed that the BPF approach addressed the problem of non-linearity in the dynamics model as well as non-Gaussian nature of the state of the NavIC satellites.

## 1.0    Introduction:

Satellite positions are essential in multiple navigation applications such as positioning, extended ephemeris technology, satellite orbit tracking and maintenance etc. Statistical orbit determination techniques can determine the satellite orbits using previously recorded broadcast ephemeris data. Orbit determination of satellite includes satellite orbit prediction with respect to time which involves solving a non-linear dynamic system governing the motion of the satellite. Thus, it is essential to address the problem of non-linearity of dynamics model and non-normality of the state space that is used to initialize the orbit prediction.

The navigation applications such as self-assisted navigation receiver needs the satellite orbits to be determined locally at the receiver using the navigation data previously available with the receiver (Stacey and Ziebart 2011). In such cases, the satellite orbit is to be determined using a single receiver data rather than from a network of multiple navigation receivers (Seppanen et al. 2011). Statistical orbit determination techniques play an important role in such situations to determine the satellite orbit using a single receiver data.

Several orbit determination algorithms have been proposed by researchers for various artificial satellites and navigation applications. Hein and Eissfeller (1997) has reported a simulation study for orbit determination of geosynchronous satellites of a European Satellite Navigation System (ENSS). This study uses Extended Kalman filter (EKF) for orbit determination of satellites using data from 8 tracking stations distributed all over the world. Carolipio et al. (2002) has reported a Kalman filter (KF) based statistical orbit determination algorithm for geostationary satellites which serve as relays to transmit messages for space-based augmentation systems (SBAS). Stacey and Ziebart (2011) and Seppanen et al. (2011) have demonstrated the Least Squares (LS) based approach for orbit determination of GPS satellites for the platform of mobile devices. Mashiku et al. (2012) have reported a Particle filter (PF) approach for orbit determination of GPS satellites using four tracking stations' data. Aghav et al. (2014) have attempted comparison between LS and EKF on the basis of a single data point for a satellite in low Earth orbit (LEO) for a prediction period of 60 sec. and has achieved km level accuracy for the orbit prediction. Shou (2014) reported a study on Unscented Kalman filter (UKF) for orbit prediction of satellites in LEO. Adaptive Kalman filter for orbit determination was demonstrated by Kavitha et al. (2015) using two-way CDMA range stations' data from India. They used data from four tracking stations across India for GSAT-10 and IRNSS-1A satellites to demonstrate the performance of algorithm. Chandrasekhar et al. (2015) have reported a numerical integration-based orbit prediction algorithm for simulated orbits of NavIC (formerly IRNSS) 1D, 1E, 1F and 1G for prediction period of 1 hr. Pardal et al. (2015) have addressed sample impoverishment problem in PF using a GPS data onboard Jason satellite. Chipade et al. (2016) have carried out a preliminary study to determine NavIC satellite orbits using LS with meter level accuracy up to 3 minutes. Shen et al. (2018) have reported ephemeris extension algorithm for BeiDou Navigation Satellite System (BDS) orbits using LS technique. They have reported the prediction of satellite positions in middle Earth orbit (MEO) and inclined geosynchronous orbit (IGSO) of BDS using precise orbit and clock products available through Multi-Global Navigation Satellite Systems (GNSS) Pilot Project (MGEX) in a post-process mode. Gamper et al. (2019) have demonstrated statistical orbit determination of resident space objects (RSO) using Ensemble Kalman filter (EnKF). These authors have also compared the performance of EnKF with UKF and observed that UKF performed better as compared to EnKF in terms of prediction accuracy. The EnKF uses a randomly chosen set of state space vectors, known as 'Ensembles' but with an assumption of Guassian distributed state space vector for satellite. Ramanathan and Chipade (2020) have developed an EKF based statistical orbit determination algorithm for NavIC satellites with meter level accuracy for short time interval of 1 hr. Chipade and Ramanathan (2021) have also demonstrated an EKF based orbit determination algorithm for GEO and GSO satellites in BeiDou navigation constellation.

Statistical orbit determination techniques such as LS, EKF UKF and EnKF made a normality assumption about the distribution of state space vector. Thus, the orbit determination algorithms based on these techniques do not address the problem of non-normality of the state space vector. Therefore, we develop a PF based statistical orbit determination algorithm for satellites in GEO and GSO of newly established NavIC (formerly known as IRNSS); an Indian regional navigation satellite system, in this paper.

NavIC constellation consists of four GSO and three GEO satellites. Three GEOs are located at 32.5$^o$ E, 83$^o$ E and 131$^o$ E, whereas the four GSOs have their longitude crossings at 55$^o$ E and 111.75$^o$ E, two in each plane (www.isac.gov.in/navigation/irnss.jsp). This constellation is designed to provide positioning services to users in India as well as around Indian subcontinent. Extended service area of NavIC lies between primary service area and area enclosed by the rectangle from latitude 30$^o$ S to 50$^o$ N and longitude 30$^o$ E to 130$^o$ E (www.isro.gov.in/irnss-programme). International GNSS Service (IGS) has its stations all over the world and one of them is BSHM, located at Haifa, Israel (32$^o$ 46' N, 35$^o$ 01' E). These IGS stations regularly upload the data to IGS data repository. We use the receiver observations for NavIC signals, collected at BSHM, Haifa in this paper.

The paper is organised as follows. The second section presents the methodology used for orbit determination. In the third section the results of orbit determination using the NavIC data has been considered. The concluding remarks are given in section four.

## 2.0    Methodology:

Let the satellite position vector be $\mathbf{r} = (x, y, z)^T$ and velocity vector be $\mathbf{v} = (v_x, v_y, v_z)^T$ known in Earth-Centred-Inertial (ECI) reference frame at some initial epoch $t_0$. Thus, the initial state space vector used for orbit prediction algorithm is $\mathbf{X} = (x, y, z, v_x, v_y, v_z)^T$. Let r and v be the magnitude of the satellite position vector and velocity vector, respectively. Let $\hat{\mathbf{X}}_k$ be the state space estimated at time $t_k$. A statistical orbit determination algorithm has two components: orbit prediction by integrating the differential equations for motion of a satellite and orbit determination using statistical methods such as LS, EKF, UKF, EnKF etc. Orbit prediction uses a set of perturbation force models that are described below.

The set of force models that determine the acceleration of the satellite due to various effects is (Ramanathan and Chipade 2020; Chipade and Ramanathan 2021)

$$\mathbf{a} = -\frac{\mu \mathbf{r}}{r^3} + \mathbf{a}_{har} + \mathbf{a}_{sun} + \mathbf{a}_{moon} + \mathbf{a}_{srp}, \qquad (1)$$

where, the Kepler's force model is

$$\mathbf{a} = -\frac{\mu \mathbf{r}}{r^3}, \tag{2}$$

with

$\mu$: Earth's Gravitational constant ($3.986004418 * 10^{14}$) (m³/s²),

$\mathbf{a}_{har}$: Acceleration due to Earth's geopotential,

$\mathbf{a}_{sun}$: Acceleration due to solar attraction,

$\mathbf{a}_{moon}$: Acceleration due to lunar attraction,

$\mathbf{a}_{srp}$: Acceleration due to solar radiation pressure,

The geopotential model of order 0 and degree 4 was considered in the present study as geopotential coefficient J2 is about 400 times larger than the next-largest value of geopotential coefficient J3 and thus for most satellite orbits reasonably good accuracy is achieved by including only J2 effect in the perturbation force model (Chobotov 2002 Chapter 9 pp. 204). The details of other acceleration forces used in this study are as given in Chipade and Ramanathan (2021). Runge-Kutta 4th order (RK4) numerical integration technique was used to propagate the satellite orbit by integrating the set of force models.

Statistical orbit determination uses knowledge of both, dynamics model that propagates satellite orbit in time and observation model. The distance from the centre of the Earth was considered as the observation model in this study in order to develop a simplified algorithm independent of any receiver observable. Thus, the element of observation vector $\mathbf{Y}_k$, at time $t_k$, can be estimated as

$$\mathbf{y}_k = \sqrt{x^2 + y^2 + z^2} \tag{3}$$

## 2.1    Least Squares (LS) Technique:

LS technique corrects the state space vector predicted using orbit propagation algorithm at time $t_k$, denoted by $\widehat{\mathbf{X}}_k^-$. Thus, the state space vector estimated using least squares as orbit determination algorithm is given by,

$$\widehat{\mathbf{X}}_k = \widehat{\mathbf{X}}_k^- + \left(\mathbf{D}^T\mathbf{D}\right)^{-1}\mathbf{D}^T\left(\mathbf{Y}_k - \mathbf{H}\widehat{\mathbf{X}}_k^-\right), \tag{4}$$

where $\mathbf{D} = \mathbf{HA}$, $\mathbf{H} = \frac{\partial \mathbf{G}}{\partial \mathbf{X}}$, $\tag{5}$

$\mathbf{G}(X,t) = [r(t)]$, $r(t) = \sqrt{(x)^2 + (y)^2 + (z)^2}$, and $\mathbf{A}$, the state transition matrix given by,

$$\mathbf{A} = \mathbf{I} + \begin{bmatrix} 0 & 0 & 0 & 1 & 0 & 0 \\ 0 & 0 & 0 & 0 & 1 & 0 \\ 0 & 0 & 0 & 0 & 0 & 1 \\ -\frac{\mu}{r^3} + \frac{3\mu x^2}{r^5} & \frac{3\mu xy}{r^5} & \frac{3\mu xz}{r^5} & 0 & 0 & 0 \\ \frac{3\mu xy}{r^5} & -\frac{\mu}{r^3} + \frac{3\mu y^2}{r^5} & \frac{3\mu yz}{r^5} & 0 & 0 & 0 \\ \frac{3\mu xz}{r^5} & \frac{3\mu yz}{r^5} & -\frac{\mu}{r^3} + \frac{3\mu z^2}{r^5} & 0 & 0 & 0 \end{bmatrix} dt. \tag{6}$$

## 2.2    Extended Kalman Filter (EKF):

For EKF, the state vector $\mathbf{X}$, is estimated under the assumption with a Gaussian distributed state vector and a linear dynamics model. EKF linearises the dynamic model using Taylor's series expansion and approximates the state vector and the observation vector with the first two terms of the Taylor series expansion. The algorithm estimates the posterior covariance associated with state estimate $\hat{\mathbf{X}}_k$; $\mathbf{P}_k$ that quantifies the uncertainty in the state estimate.

At time $t_k$, the state vector and the error covariance can be computed using the state transition matrix $\mathbf{A}$. Thus,

$$\hat{\mathbf{X}}_k^- = \mathbf{A}\hat{\mathbf{X}}_{k-1},$$

$$\mathbf{P}_k^- = \mathbf{A}\mathbf{P}_{k-1}\mathbf{A}^T + \mathbf{Q}, \tag{7}$$

where, $\mathbf{Q}$ is the diagonal, process noise covariance matrix.

The Kalman gain $\mathbf{K}$ is computed from the observation model using the $\mathbf{H}$-matrix as

$$\mathbf{K} = \mathbf{P}_k^-\mathbf{H}^T(\mathbf{H}\mathbf{P}_k^-\mathbf{H}^T + \mathbf{R})^{-1}, \tag{8}$$

where $\mathbf{R}$ is the measurement covariance and matrix $\mathbf{H}$ is given by equation (5). The state estimate is updated using the Kalman gain and the innovation term $(\mathbf{Y}_k - \mathbf{H}\hat{\mathbf{X}}_k^-)$, where $\mathbf{Y}_k$ was the observed satellite range at time $t_k$. Thus,

$$\hat{\mathbf{X}}_k = \hat{\mathbf{X}}_k^- + \mathbf{K}(\mathbf{Y}_k - \mathbf{H}\hat{\mathbf{X}}_k^-),$$

$$\mathbf{P}_k = [\mathbf{I} - \mathbf{K}\mathbf{H}]\mathbf{P}_k^-. \tag{9}$$

The algorithm was initialised with a state vector $\hat{\mathbf{X}}_0$ computed based on previously known orbital parameters in Earth-Centred Inertial (ECI) reference frame and an a priori covariance associated with state estimate, $\mathbf{P}_0$; for statistical orbit determination of satellites.

## 2.3    Unscented Kalman Filter (UKF):

EKF uses only the first and second order terms of the Taylor series expansion and neglects the higher order terms. The UKF algorithm can be applied to non-linear Gaussian orbit determination problem. The UKF follows the intuition that, with a fixed number of parameters, it should be easier to approximate a Gaussian distribution than an arbitrary, non-linear function (Turner and Sherlock 2013). The UKF assumes that the satellite state space vector is Gaussian. The orbit propagation was initialised with the state space vector at an initial time $t_0$, $\mathbf{X}_0$; with the associated mean vector and covariance matrix, given as

$$\hat{\mathbf{X}}_0 = E[\mathbf{X}_0] \text{ and} \tag{10}$$

$$\mathbf{P}_0 = E\left[\left(\mathbf{X}_0 - \hat{\mathbf{X}}_0\right)\left(\mathbf{X}_0 - \hat{\mathbf{X}}_0\right)^T\right] \tag{11}$$

respectively. The state space vector is augmented with the process and measurement noise terms to generate the sigma points. The superscript $a$ stands for the augmented vector or matrix. The mean and covariance matrix associated with augmentation at an initial time $t_0$ are

$$\hat{\mathbf{X}}_0^a = E[\hat{\mathbf{X}}_0^a] = [\hat{\mathbf{X}}_0^T \quad \mathbf{0} \quad \mathbf{0}], \text{ and} \tag{12}$$

$$\mathbf{P}_0^a = E\left[\left(\mathbf{X}_0^a - \hat{\mathbf{X}}_0^a\right)\left(\mathbf{X}_0^a - \hat{\mathbf{X}}_0^a\right)^T\right]. \tag{13}$$

The initial weights associated with the initial mean state and augmentation to compute sigma points were respectively,

$$W_0^{(mean)} = \lambda/(L + \lambda) \text{ and,} \tag{14}$$

$$W_0^{(sigma-points)} = \lambda/(L + \lambda) + (1 - \alpha^2 + \beta). \tag{15}$$

Here, $\beta$ is the prior knowledge or information term with a value of 2 for a Gaussian initial distribution and $\alpha = 10^{-3}$. The $\lambda$ is a scaling parameter details about which can be found in Mashiku (2013). The dimension of the state space vector denoted by L was 6 in this study.

The sigma points were computed using the mean state, covariance matrix associated with the augmentation of the state space vector with process and observation noise and the corresponding initial weights. Thus, the expression that generates sigma points at any time $t_{k-1}$, is given by

$$\boldsymbol{\mathcal{X}}_{k-1}^a = \left[\hat{\mathbf{X}}_0^a \quad \hat{\mathbf{X}}_0^a \pm \left(\sqrt{(L + \lambda)\mathbf{P}_{k-1}^a}\right)\right]. \tag{16}$$

The matrix of sigma points, $\boldsymbol{\mathcal{X}}_{k-1}^a$; has 2L+1 columns.

The weights associated with sigma points for time t > 0, i.e., at any time $t_{k-1}$ are given by,

$$W_{k-1} = W_{k-1}^{(mean)} = W_{k-1}^{(sigma-points)} = 1/2(L + \lambda). \tag{17}$$

The sigma points were propagated through non-linear dynamics model from time $t_{k-1}$ to $t_k$, to obtain the a-priori state estimate. The observations corresponding to the sigma points were propagated in time using the observation model. Thus, time updated sigma points and observations were computed as follows.

$$\boldsymbol{\mathcal{X}}_k^a = \mathbf{f}(\boldsymbol{\mathcal{X}}_{k-1}^a) = \mathbf{A}\boldsymbol{\mathcal{X}}_{k-1}^a, \text{ and} \tag{18}$$

$$\boldsymbol{\mathcal{Y}}_k^a = \mathbf{H}\boldsymbol{\mathcal{X}}_{k-1}^a, \tag{19}$$

where $\mathbf{A}$ is the state transition matrix given in (6), $\boldsymbol{\mathcal{Y}}_k^a$, is the observation vector associated with sigma points. The sigma points thus obtained using (18) were the unscented transformed sigma points.

Thus, a-priori state estimate $\widehat{\mathbf{X}}_k^-$ and the corresponding associated observation update $\widehat{\mathbf{Y}}_k^-$ were obtained as the weighted sum,

$$\widehat{\mathbf{X}}_k^- = \sum_{i=0}^{2L} W_i \, \boldsymbol{\mathcal{X}}_{i,k}^a, \text{ and} \tag{20}$$

$$\widehat{\mathbf{Y}}_k^- = \sum_{i=0}^{2L} W_i \, \boldsymbol{\mathcal{Y}}_{i,k}^a. \tag{21}$$

The covariance matrix of predicted errors in a priori state estimate $\widehat{\mathbf{X}}_k^-$, is given by,

$$\mathbf{P}_k^- = \sum_{i=0}^{2L} W_i \left[ \boldsymbol{\mathcal{X}}_{i,k}^a - \widehat{\mathbf{X}}_k^- \right] \left[ \boldsymbol{\mathcal{X}}_{i,k}^a - \widehat{\mathbf{X}}_k^- \right]^T. \tag{22}$$

The state estimate was computed in the same format as innovation term in EKF. A sigma-point gain was used in a similar manner as in EKF to weigh the innovation term to obtain the state estimate. The sigma point gain was computed as follows:

$$\boldsymbol{\mathcal{K}} = \mathbf{P}_{\mathbf{X}_k \mathbf{Y}_k} \mathbf{P}_{\mathbf{Y}_k \mathbf{Y}_k}^{-1}, \tag{23}$$

where,

$$\mathbf{P}_{\mathbf{X}_k \mathbf{Y}_k} = \sum_{i=0}^{2L} W_i \left[ \boldsymbol{\mathcal{X}}_k^a - \widehat{\mathbf{X}}_k^- \right] \left[ \boldsymbol{\mathcal{Y}}_k^a - \widehat{\mathbf{Y}}_k^- \right]^T, \tag{24}$$

and

$$\mathbf{P}_{\mathbf{Y}_k \mathbf{Y}_k} = \sum_{i=0}^{2L} W_i \left[ \boldsymbol{\mathcal{Y}}_k^a - \widehat{\mathbf{Y}}_k^- \right] \left[ \boldsymbol{\mathcal{Y}}_k^a - \widehat{\mathbf{Y}}_k^- \right]^T. \tag{25}$$

Thus, the state estimate and associated covariance matrix were updated as

$$\widehat{\mathbf{X}}_k = \widehat{\mathbf{X}}_k^- + \boldsymbol{\mathcal{K}}(\mathbf{Y}_k - \widehat{\mathbf{Y}}_k^-), \tag{26}$$

$$\mathbf{P}_k = \mathbf{P}_k^- - \boldsymbol{\mathcal{K}} \mathbf{P}_{\mathbf{Y}_k \mathbf{Y}_k} \boldsymbol{\mathcal{K}}^T. \tag{27}$$

The UKF was used to estimate the satellite orbits of NavIC satellites, results of which are discussed in the subsequent section.

### 2.4    Ensemble Kalman Filter (EnKF):

Evensen (1994) proposed EnKF to address the difficulties in using EKF. EnKF generates a random sample from initial Gaussian distribution of the state space vector termed as 'Ensemble'. This ensemble is randomly generated only once and is used to propagate the state of the satellite orbit in time unlike UKF, where sigma points are generated with every state estimate at every time step. However, EnKF assumes state vector distribution and errors associated with the dynamics and observation models to be Gaussian.

The EnKF was initialised with the satellite state vector at time $t_0$, $\mathbf{X}_0$; with the associated mean vector $\widehat{\mathbf{X}}_0$ and covariance matrix $\mathbf{P}_0$ given by (10) and (11) respectively. A random sample of size N was generated which is called as ensemble of size N using the initial probability distribution of the state vector. Let $\left\{\mathbf{X}_k^{(i)}\right\}_{i=1}^N$ be the ensemble of size N randomly drawn from multivariate normal (MVN) distribution with mean $\widehat{\mathbf{X}}_0$ and variance $\mathbf{P}_0$. Each of the ensemble member was propagated in time using the dynamics model. The mean and the variance of the propagated ensemble matrix at time $t_k$ were computed as

$$\overline{\mathbf{X}}_k = \frac{1}{N}\sum_{i=1}^N \mathbf{X}_k^{(i)}, \tag{28}$$

$$\overline{\mathbf{P}}_k = \frac{\gamma}{N-1}\mathbf{P}_0, \tag{29}$$

where $\gamma$ is a scaling parameter used to inflate the underestimated variance matrix of the ensemble in order to avoid the divergence of the EnKF. Gamper et al. (2019) have studied the optimization of inflation parameter and have estimated the value of $\gamma$ to be 0.95 as the best value with minimum prediction error. The same value of $\gamma$ was used in the present study. The observation vector associated with ensemble mean was computed as in (21) with $W_i = \frac{1}{N}$. The ensemble Kalman gain, $\boldsymbol{\mathcal{K}}^e$ was estimated as

$$\boldsymbol{\mathcal{K}}^e = \mathbf{P}_{\mathbf{X}_k\mathbf{Y}_k}\mathbf{P}_{\mathbf{Y}_k\mathbf{Y}_k}^{-1}, \tag{30}$$

where the covariance and variance matrices, $\mathbf{P}_{\mathbf{X}_k\mathbf{Y}_k}$ and $\mathbf{P}_{\mathbf{Y}_k\mathbf{Y}_k}$ were estimated as in (24) and (25) respectively with $W_i = \frac{1}{N-1}$ for i = 1,2…N.

The EnKF state estimate and corresponding variance matrix were estimated as

$$\widehat{\mathbf{X}}_k = \overline{\mathbf{X}}_k + \boldsymbol{\mathcal{K}}^e\left(\mathbf{Y}_k - \widehat{\mathbf{Y}}_k^-\right), \text{ and} \tag{31}$$

$$\mathbf{P}_k = \overline{\mathbf{P}}_k - \boldsymbol{\mathcal{K}}^e\mathbf{P}_{\mathbf{Y}_k\mathbf{Y}_k}\boldsymbol{\mathcal{K}}^{e^T}. \tag{32}$$

More theoretical details about the EnKF may be found in Gamper et al. (2019).

## 2.5    Particle Filter (PF):

The EKF algorithm addressed the problem of non-linear dynamics model, but assumed the Gaussian distribution for the state space vector. UKF is based on the principle of approximating a Gaussian distribution with a set of sigma points generated using the mean and covariance of process noise and observation errors (Merwe et al. 2000, Turner and Sherlock 2013). EnKF uses a randomly generated set of state space vectors which are assumed to be Gaussian distributed to determine the satellite orbit (Gamper et al. 2019). However, the state space vector of the satellite orbit need not be Gaussian distributed. Thus, a technique of addressing the non-linearity of the dynamics model and the non-Gaussian state space vector is essential for the orbit determination. The PF addresses both these issues. The PF is a sequential Bayesian algorithm that can be used for nonlinear systems and non-Gaussian distributed states. The central idea of a PF is to represent the required probability density function (PDF) of the satellite state vector by a set of $N \gg 1$ random samples (particles) $\left\{ \mathbf{X}_k^{(i)} \right\}_{i=1}^N$ drawn from the PDF with a Gaussian distribution assumption.

We develop the algorithm of orbit determination using the BPF as follows (Mashiku et al. 2012, Pardal et al. 2015).

<u>Step 1: Initialization</u> at $k = 0$, for $i = 1, ..., N$

Sample N particles, $\mathbf{X}_0^{(i)}$, for $i = 1, ..., N$ from the prior $\text{MVN}(\hat{\mathbf{X}}_0, \mathbf{P}_0)$. The constant variance covariance matrix associated with state space vector was assumed to be $\mathbf{P}_0$. Calculate the weights associated with particles using initial Gaussian distribution as

$$w_0^i = P\big[\mathbf{Y_0}|\mathbf{X}_0^{(i)}\big], \text{ where } \mathbf{Y_0} \sim \text{MVN}(\mathbf{y_0}, \mathbf{R}). \tag{33}$$

Normalise the weights as $w_0^i = \frac{w_0^i}{w_T}$ where $w_T = \sum_i^N w_0^i$.

<u>Step 2: Time Update or Orbit Prediction:</u>

The randomly generated particles were updated in time $t_k$ for $k \geq 1$ and for $i = 1, ..., N$ through dynamics model by numerically integrating the particles using 4$^{th}$ order Runge-Kutta method. Thus,

$$\mathbf{X}_k^{(i)} = \int_{t_0}^{t_k} f(\mathbf{X}_0^{(i)}) d\mathbf{X}_{k-1} \text{ with } f(\mathbf{X}_0^{(i)}) = \dot{\mathbf{X}}_0^{(i)}. \tag{34}$$

<u>Step 3: Measurement Update:</u>

Compute the relative likelihood $q_k^i$ as

$$q_k^i = P\big[\mathbf{Y}_k = \mathbf{y}_k|\mathbf{X}_k^{(i)}\big] \tag{35}$$

Normalise the relative likelihood values as $q_k^i = \frac{q_k^i}{q_T}$ where $q_T = \sum_i^N q_k^i$.

The weights were updated under the assumption that importance distribution in this case is also equal to the prior density and thus,

$$w_k^i = w_{k-1}^i P\big[\mathbf{Y}_k|\mathbf{X}_k^{(i)}\big], \text{ where } \mathbf{Y}_k \sim \text{MVN}(\mathbf{y}_k, \mathbf{R}). \tag{36}$$

Normalise the weights as $w_k^i = \frac{w_k^i}{w_T}$ where $w_T = \sum_i^N w_k^i$.

<u>Step 4: Resampling:</u>

Particle filter suffers the problem of sample degeneracy and thus resampling was performed to address sample degeneracy. In the resampling step, we generate a sample of size N proportional to relative likelihood values $q_k^i$. This resample was the set of particles distributed according to $P[\mathbf{X}_k|\mathbf{Y}_k]$.

<u>Step 5: Roughening:</u>

Roughening is a way to tackle the sample impoverishment by adding a random noise to the resampled particles and thus producing a diverse set of resampled particles. In this step, a posteriori particles were modified after the resampling step, given by;

$$\widehat{\mathbf{X}_k^{*(i)}}(m) = \widehat{\mathbf{X}_k^{(i)}}(m) + \Delta\mathbf{X}(\mathbf{m}), \text{m} = 1, 2...\text{n; with} \tag{37}$$

$$\Delta\mathbf{X}(\mathbf{m}) \sim \mathbf{N}(\mathbf{0}, \mathrm{K}\mathbf{M}(\mathbf{m})\mathrm{N}^{-1/n}), \tag{38}$$

where $\Delta\mathbf{X}(\mathbf{m})$ is a zero mean Gaussian random variable, K is the tuning parameter, N is the number of particles, n is the state space dimension (which is 6 in this problem) and $\mathbf{M}$ is a vector of the maximum difference between the particle elements before roughening. The $\mathrm{m}^{\mathrm{th}}$ element of the $\mathbf{M}$ vector is given as

$$\mathbf{M}(m) = {}^{max}_{i,j}|\widehat{\mathbf{X}_k^{(i)}}(m) - \widehat{\mathbf{X}_k^{(j)}}(m)|, \text{m=1, 2...n.} \tag{39}$$

The value of the tuning parameter K, was selected as 0.1 (Pardal et al. 2015).

The state estimate was then computed as

$$\widehat{\mathbf{X}}_k = \text{avg}\left\{\widehat{\mathbf{X}_k^{*(i)}}, \text{ for i=1,..., N}\right\}. \tag{40}$$

The scheme thus developed to estimate state space as given in (40) is known as Bootstrap Particle Filter (BPF). We have developed this approach of BPF implementation for the orbit determination problem.

As stated earlier, the solution to overcome sampling degeneracy is to resample the particles. However, the choice of number of particles to be generated to estimate the satellite position using BPF is important. Let $\mathrm{N}_{\mathrm{eff}}$ denote the effective sample size, which is estimated as, in Arulampalam et al. (2002) and Mashiku et al. (2012) as

$$\mathrm{N}_{\mathrm{eff}} = \frac{1}{\sum_{i=1}^N (w_k^{(i)})^2}. \tag{41}$$

Also let $N_{th}$ denote a lower threshold for the effective sample size, which is arbitrarily chosen with respect to the accuracy desired (Mashiku et al. 2012). The larger the threshold, $N_{th}$, the accurate the PDF results and more the computational cost incurred. The lower the value of $N_{eff}$, severe the problem of sampling degeneracy. Thus, choice of the value of $N_{eff}$ is very important. In the context of the problem at hand, balancing the computational time and the model complexity threshold $N_{th}$, was set at 10. Using the weights associated with the particles of size $N_{th} = 10$, the $N_{eff}$ was computed at the initiation of the BPF. The integral values of $N_{eff}$ thus computed using the generated particles and their weights, were observed to vary between 5 to 15. The particle size of 5 is very low to generate the prior and causes sample degeneracy severely. The particle size of 15 resulted in extensive computational time for processing the data and estimating the satellite position. Thus, $N_{eff}$ was selected as ten and particles of size ten were generated. The ensemble size was selected as ten in order to compare the performance of EnKF with BPF. The details about the prediction accuracy criterion and residual analysis are presented in the subsequent sections.

## 2.6 Criterion Used for the Evaluation of Predicted Orbit Accuracy:

The orbit determination algorithm was initialised with satellite coordinates computed in ECI reference frame. However, the ranging errors primarily come from the radial direction of satellite orbit. Thus, satellite coordinates predicted in ECI reference frame were transformed to radial (R), along track (S), across track (W) (RSW) reference frame and the prediction error in the radial direction of satellite was analysed in the present study. The root mean square error (RMS error or RMSE) was used in the present study to evaluate the orbit accuracy. The predicted satellite coordinates using the statistical orbit determination algorithms were compared with satellite coordinates estimated using the real navigation data that was collected at the IGS station. RMS error was computed as

$$\text{RMSE} = \sqrt{\frac{\sum_{i=1}^{T}\left(Q_{i,\text{pred}} - Q_{i,\text{act}}\right)^2}{T}}, \tag{42}$$

where T is the total duration over which the orbit is predicted. Here, T=3600, as the orbit predicted for next one hour at an interval of 24 sec. The quantities $Q_{\text{pred}}$ and $Q_{\text{act}}$ are the predicted and actually estimated satellite coordinates in radial direction, respectively.

## 2.7 Henze-Zirkler Multivariate Normality Test:

The statistical orbit determination algorithms such as EKF and UKF assume the Gaussian distributed state space vector of a satellite orbit. However, this may not be true with the real time satellite orbit data. Thus, it is important to examine the residuals of predicted satellite orbits for their normality. The state space vector being a 6-dimensional vector, a multivariate normality test was needed to examine the normality of the residuals obtained.

Thode (2002 Chap. 9 pp. 220) recommends the multivariate normality test proposed by Henze and Zirkler (1990), as it has better power properties over many other alternatives. Henze-Zirkler (HZ) test is based on the weighted integral of the difference between the empirical characteristic function (ECF) and its pointwise limit. The hypothesis to be tested is

$H_0$: The random vector (state space residual vector $(\mathbf{X} - \hat{\mathbf{X}})$ in the context of OD problem) has multivariate normal distribution with mean zero and covariance matrix $\mathbf{Q}$.

The test statistics $T_\beta$ is defined as

$$T_\beta = n\left(4I_{(\mathbf{S}\ \text{Singular})} + D_{m,\beta}I_{(\mathbf{S}\ \text{Non-Singular})}\right), \tag{43}$$

where, $\mathbf{S}$ is the sample covariance matrix which is an estimate of covariance matrix $\mathbf{Q}$, I is the indicator function, m is sample size and $D_{m,\beta}$ is given by,

$$D_{m,\beta} = \frac{1}{m^2}\sum_{j,k=1}^{m} exp\left(-\frac{\beta^2}{2}\left\|\boldsymbol{\Theta}_j - \boldsymbol{\Theta}_k\right\|^2\right) + (1+2\beta^2)^{-n/2} - \frac{2(1+\beta^2)^{-n/2}}{m}\sum_{j=1}^{m} exp\left(-\frac{\beta^2}{2(1+\beta^2)}\boldsymbol{\Theta}_j^2\right), \tag{44}$$

where

$$\left\|\boldsymbol{\Theta}_j - \boldsymbol{\Theta}_k\right\|^2 = \left(\mathbf{X}_j - \mathbf{X}_k\right)^T \mathbf{S}^{-1}\left(\mathbf{X}_j - \mathbf{X}_k\right). \tag{45}$$

The weighting function, $\beta$ can be optimally chosen as,

$$\beta = \frac{1}{\sqrt{2}}\left(\frac{m(2n+1)}{4}\right)^{1/(n+4)}, \tag{46}$$

where n is the dimension of the state space vector and m is the sample size (see Thode 2002, Chap. 9 pp. 215).

The residuals of predicted satellite orbit were computed in ECI reference frame before computation of prediction accuracy in radial direction and were tested for normality. A Matlab function `HZmvntest` developed by Trujillo-Ortiz et al. (2007) was used in this study to test the normality of residuals. The detailed results of the residual analysis are presented in the next section.

### 3.0    Results and Discussion:

We discuss the results related to the prediction of satellite coordinates in radial direction for NavIC GEO and GSO satellites using BPF and compare them with the traditional methods of orbit determination such as LS, EKF, UKF and EnKF. The diagonal elements of a priori covariance matrix ($\mathbf{P}_0$) associated with the state vector were initialised as (10 m, 10 m, 10 m, 0.1 m/s, 0.1 m/s, 0.1 m/s). The unit measurement covariance ($\mathbf{R}$) was considered, whereas the diagonal elements of process noise covariance ($\mathbf{Q}$) were initialised at (0.001 m, 0.001 m, 0.001 m, 1e-6 m/s, 1e-6 m/s, 1e-6 m/s) (Ramanathan and Chipade 2020). The NavIC satellite orbits were predicted for short-term duration of 1 hr. at time step of 24 sec. using statistical orbit determination algorithms such as LS, EKF, UKF, EnKF and BPF consecutively for a period of five days. The navigation data collected for five days during 1 March 2020 to 5 March 2020 at IGS station BSHM, Haifa was used to initialise the statistical orbit determination algorithms. The satellite coordinates were computed and predicted in ECI reference frame and were later transformed to RSW frame to compute the prediction error in radial direction of NavIC satellites.

Table 1 shows the RMSE in radial direction of NavIC satellites in GSO and GEO orbits after 1 hr., predicted using LS, EKF, UKF, EnKF and BPF. It can be observed that the RMSE decreased from 85.09 m to 25.49 m in radial direction of NavIC GSO satellite when the orbit determination technique of LS was enhanced with BPF to address the problems of non-linear dynamics model and non-Gaussian state space vector. It was also observed that the RMSE decreased from 93.82 m to 34.79 m in radial direction of NavIC GEO satellite when algorithm of LS was changed to BPF. Table 1 clearly showed that performance of EKF was better as compared to LS and performance of BPF was better as compared to LS and EKF both, in terms of RMSE. The performance of EnKF in terms of RMSE was observed to be worse than UKF and BPF on all of the days except for the GEO satellite on the last day (Table 1). In case of NavIC GEO satellite the RMSE in radial direction, predicted using UKF and BPF was observed to have no significant difference; whereas in case of NavIC GSO satellite RMSE in radial direction, predicted using UKF was observed to be less as compared to that predicted using BPF. However, the difference in RMSE due to UKF and BPF was not more than 3 m. This observation made an impression that UKF performed better as compared to BPF for NavIC satellite orbit determination for short duration which assumed a Gaussian state space vector. The Gaussian assumption of the state space vector was tested using the residuals with the HZ test for normality. The HZ test for normality was not performed in case of LS technique as significantly high RMSE values were observed as compared to UKF and BPF. The BPF method addresses the orbit determination problem for non-Gaussian state space non-linear dynamics model. Thus, HZ test for normality was not performed for residuals due to BPF.

Table 2 shows the test statistic and p-values of the HZ test for multivariate normality for NavIC GSO and GEO satellites; predicted using EKF UKF and EnKF for the period of 5 days. The p-value < 0.05 (a non-zero digit appeared only after the 6[th] decimal place in the p-value) showed that the null hypothesis of the normality of the state space vector was rejected indicating non-Gaussian state space vector. Thus, the assumption for Gaussian state space vector in EKF, UKF and EnKF algorithms was violated indicating the need for the method that address the non-Gaussian state space vector.

Table 1: RMSE in radial direction of NavIC satellites in GSO and GEO using LS, EKF, UKF, EnKF and BPF (in meter) (T = 1 hr.)

| Day | LS | | EKF | | UKF | | EnKF | | BPF | |
|---|---|---|---|---|---|---|---|---|---|---|
| | GSO | GEO | GSO | GEO | GSO | GEO | GSO | GEO | GSO | GEO |
| 1 Mar. 2020 | 62.84 | 89.27 | 53.22 | 81.11 | 3.73 | 28.80 | 6.52 | 33.20 | 6.22 | 28.93 |
| 2 Mar. 2020 | 73.24 | 93.82 | 64.28 | 85.87 | 10.32 | 34.20 | 15.62 | 39.20 | 11.39 | 34.79 |
| 3 Mar. 2020 | 79.44 | 90.19 | 70.83 | 82.14 | 17.65 | 30.51 | 22.99 | 34.76 | 18.42 | 30.49 |
| 4 Mar. 2020 | 85.09 | 75.40 | 76.44 | 66.70 | 24.18 | 13.98 | 28.97 | 19.19 | 25.49 | 14.61 |
| 5 Mar. 2020 | 80.89 | 55.28 | 72.39 | 45.19 | 19.71 | 11.27 | 24.56 | 8.13 | 20.08 | 12.11 |

Table 2: Test statistic (TS) and p-value (p) of HZ test for NavIC GSO and GEO satellites predicted using EKF, UKF and EnKF (T = 1 hr.)

| Day | GSO | | | GEO | | |
|---|---|---|---|---|---|---|
| | EKF | UKF | EnKF | EKF | UKF | EnKF |
| | TS (p) | TS (p) | TS (p) | TS (p) | TS (p) | TS (p) |
| 1 Mar. 2020 | 153.18 (0.00) | 130.57 (0.00) | 72.89 (0.00) | 149.70 (0.00) | 89.67 (0.00) | 68.03 (0.00) |
| 2 Mar. 2020 | 153.39 (0.00) | 125.67 (0.00) | 63.07 (0.00) | 110.06 (0.00) | 90.50 (0.00) | 56.64 (0.00) |
| 3 Mar. 2020 | 122.62 (0.00) | 132.14 (0.00) | 66.87 (0.00) | 87.46 (0.00) | 87.18 (0.00) | 47.55 (0.00) |
| 4 Mar. 2020 | 154.30 (0.00) | 155.33 (0.00) | 75.65 (0.00) | 93.42 (0.00) | 79.71 (0.00) | 47.52 (0.00) |
| 5 Mar. 2020 | 155.60 (0.00) | 150.91 (0.00) | 77.14 (0.00) | 106.35 (0.00) | 81.36 (0.00) | 54.15 (0.00) |

Figure 1 shows the variation in the residuals in radial direction of NavIC GSO and GEO satellites using UKF. The residual errors were observed to be gradually increasing over time with a step change at every 15 min. This variation observed in the residuals in radial direction of NavIC satellites predicted using UKF indicates that UKF was not suitable as an orbit determination algorithm for non-Gaussian state space vector though it showed comparatively less RMSE.

Figure 2 shows the variation in residuals in radial direction of NavIC satellites predicted using BPF over 5 days. The residuals showed a random distribution over a period of 1 hr. without any step changes as were seen in residuals due to UKF predictions. This indicates that the BPF have addressed the problem of non-normality of the state space data by generating a number of particles. This was reflected in these residual plots showing randomly distributed errors without deviating farther from the true value of the state estimate.

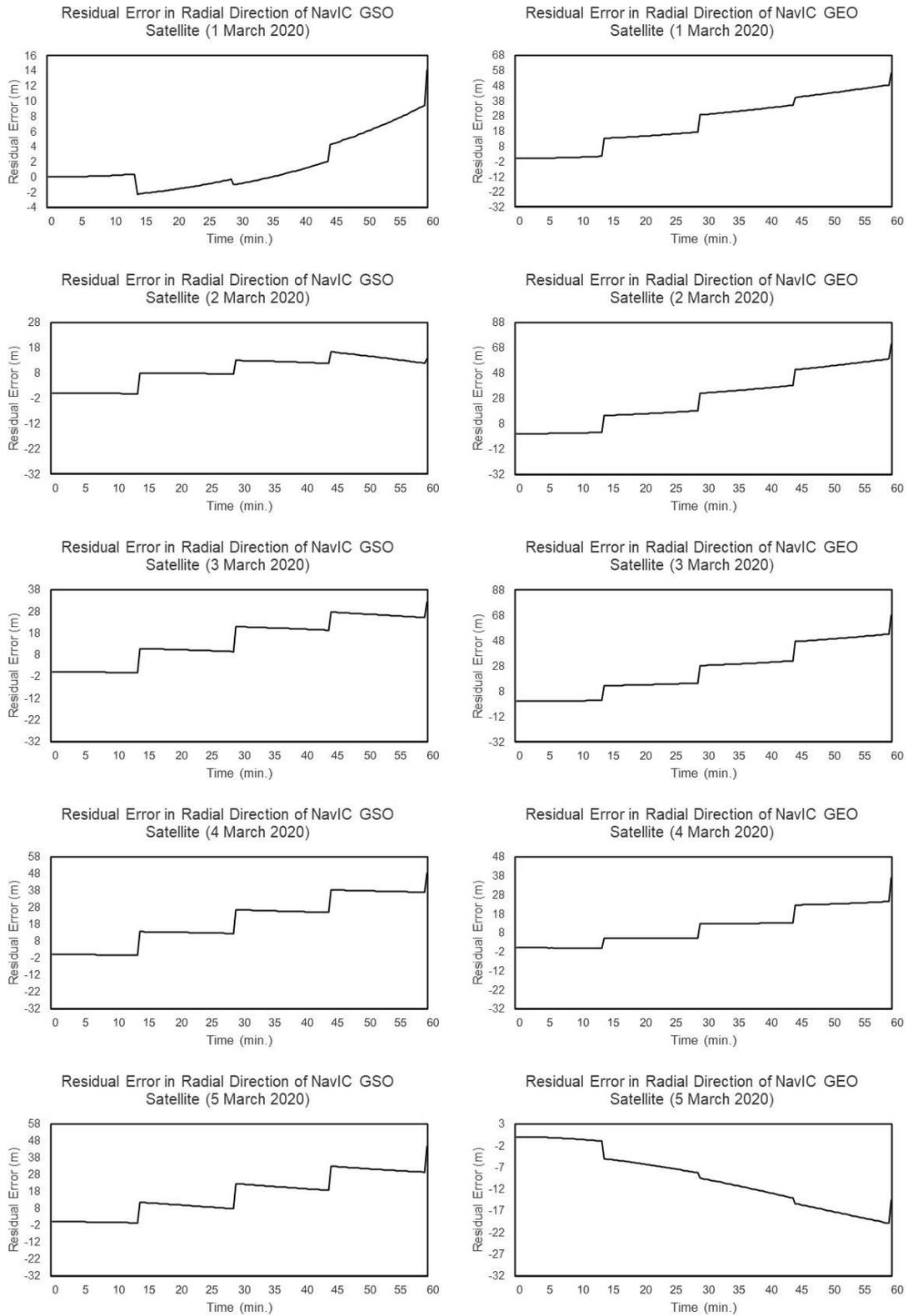

Figure 1: Residual Error in Radial Direction of NavIC satellites predicted using UKF

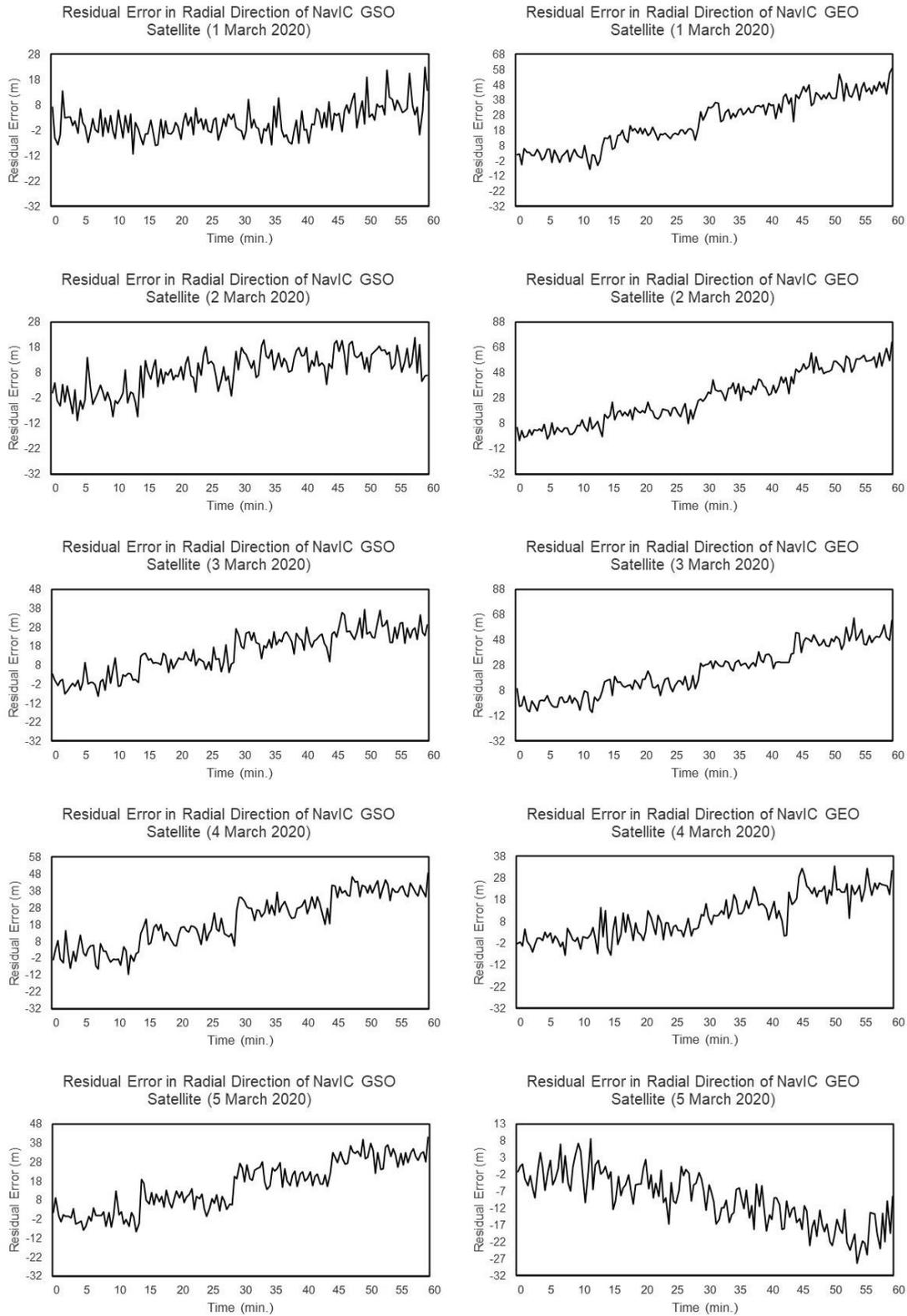

Figure 2: Residual Error in Radial Direction of NavIC satellites predicted using BPF

BPF achieved meter level accuracy in the radial direction of NavIC satellites. The performance of BPF was found to be best when compared with LS, EKF, UKF and EnKF algorithms for orbit determination in terms of RMSE and random behaviour of residuals. BPF with particle size of 10 also balanced the model complexity and computational time for orbit prediction of satellites. Thus, BPF was found to be the most suitable algorithm for statistical orbit determination of NavIC GSO and GEO satellites.

## 4.0    Conclusions:

The satellite orbit determination is essential for various navigation applications. In this paper, the statistical orbit determination algorithms were studied and evaluated for their performance for newly established satellites in NavIC constellation. The satellite orbits in NavIC constellation were determined using LS, EKF, UKF, EnKF and BPF algorithms. EKF is the traditional algorithm that is used extensively for orbit determination. However, it assumes the Gaussian state space vector and linearises the dynamics model for orbit determination. UKF and EnKF also assume the Gaussian state space vector for orbit determination and thus often are not suitable theoretically for orbit determination. Thus, BPF is the most suitable statistical orbit determination algorithm that can address the difficulties posed by traditional algorithms. We use this algorithm and compare the orbit determination performance of NavIC satellites with some of the existing algorithms. In case of NavIC satellites, BPF addressed the non-linearity and non-Gaussianity of the state of the model by generating particles around the mean value of the state at the start of the algorithm. Meter level prediction accuracy was achieved using BPF up to 1 hr. of predictions. Random behaviour of the residuals due to BPF indicates that the problem of non-Gaussian state of the satellite was addressed using BPF. BPF outperformed in terms of prediction accuracy as compared to traditional methods such as LS, EKF, UKF and EnKF for short-term prediction of satellite orbit. The study has established BPF as the most suitable statistical orbit determination algorithm that can be utilised in future for navigation applications such as Self-Assisted NavIC. The study has scope to improve the prediction accuracy of the BPF with addition of clock parameters in the state space vector and utilising the code phase and carrier phase observables from a navigation receiver.

## 5.0    Acknowledgements:

The first author would like to acknowledge the Science and Engineering Research Board (SERB) for the partial support through the grant MTR/2020/000126 (Mathematical Research Impact Centric Support (MATRICS)). The second author's work is a part of the Ph.D. research carried out at the Department of Statistics, Savitribai Phule Pune University, Pune, India. The second author is thankful to Shri. N. M. Desai, Director, Space Applications Centre (SAC), ISRO, Ahmedabad, India for providing the opportunity to carry out this work. She is also grateful to Dr. M. R. Pandya, Head, AED/EPSA/SAC; Dr. B. K. Bhattacharya, Group Director/BPSG/EPSA/SAC; Dr. I. M. Bahuguna, Dy. Director/EPSA for their constant support and encouragement.

## 6.0    References:

Aghav S and Gangal S (2014) Simplified Orbit Determination Algorithm for Low Earth Orbit Satellites using Spaceborne GPS Navigation Sensor. Artificial Satellites 49(2): 81-99.

Arulampalam MS, Maskell N, Gordon, and Clapp T (2002) A Tutorial on Particle Filters for Online Nonlinear/non-Gaussian Bayesian Tracking. IEEE Transactions on Signal Processing 50(2): 174–188.

Carolipio EM, Pandya NK, Grewal MS (2002) GEO Orbit Determination via Covariance Analysis with a Known Clock Error. Navigation 48(4): 255-260.

Chandrasekhar MV, Rajarajan D, Satyanarayana G, Tirmal N, Rathnakara SC, Ganeshan AS (2015) Modernized IRNSS Broadcast Ephemeris Parameters. Control Theory and Informatics 5(2): 1-9.

Chipade RA, Shukla AK, Shukla AP (2016) Self-Assisted Ephemeris in IRNSS User Receiver. Internal Report of SAC/ISRO (Restricted Distribution) TDP Close Out Report. SAC/TDP/2016/E120.

Chipade RA and Ramanathan TV (2021) Extended Kalman Filter based Statistical Orbit Determination for Geostationary and Geosynchronous Satellite Orbits in BeiDou Constellation. Contributions to Geophysics and Geodesy 51(1): 25-46.

Chobotov VA (2002) Orbital Mechanics, 3rd Ed., AIAA, Virginia.

Evensen G (1994) Sequential Data Assimilation with a Non-Linear Quasi-Geostrophic Model using Monte Carlo Methods to Forecast Error Statistics. Journal of Geophysical Research 99(C5): 10143-10162.

Gamper E, Kebschull C, Stoll E (2019) Statistical Orbit Determination using The Ensemble Kalman Filter. Proc. 1st NEO and Debris Detection Conf., Darmstadt, Germany. (http://neo-sst-conference.sdo.esoc.esa.int).

Hein GW and Eissfeller B (1997) Orbit Determination of Geosynchronous Satellites of a European Satellite Navigation System (ENSS). Proc. 12th Int. Sym. Space Flight Dynamics: 59-64.

Henze N and Zirkler B (1990) A Class of Invariant Consistent Tests for Multivariate Normality. Communications in Statistics-Theory and Methods 19: 3595-3617.

Kavitha S, Mula P, Babu R, Ratnakara SC, Ganeshan AS (2015) Adaptive Extended Kalman filter for Orbit Estimation of GEO Satellites. J. Environment and Earth Science 5(3): 1-10.

Mashiku AK, Garrison JL, Carpenter JR (2012) Statistical Orbit Determination using the Particle filter for incorporating Non-Gaussian Uncertainties. AIAA/AAS Astrodynamics Specialist Conference - August 2012: 1-12.

Mashiku AK (2013) Information Measures for Statistical Orbit Determination. Ph. D. Dissertation, Purdue University, Indiana.

Merwe R, Doucet A, de Freitas N, and Wan E. (2000) The Unscented Particle Filter. Technical report, Cambridge University Engineering Department.

Pardal PCPM, Kuga HK, de Moraes RV (2015) The Particle Filter Sample Impoverishment Problem in the Orbit Determination Application. Hindawi Mathematical Problems in Engineering Article ID 1680459. doi:10.1155/2015/168045.

Ramanathan TV and Chipade RA (2020) Statistical Orbit Determination Algorithm for Satellites in Indian Navigation Constellation (NavIC): Towards Extended Ephemeris Technology for NavIC Receiver. Artificial Satellites 55(2): 29-40.

Seppanen M, Perala T, Piche R (2011) Autonomous Satellite Orbit Prediction. Proceedings of the 2011 International Technical Meeting of The Institute of Navigation 554-564.

Shen Z, Peng J, Liu W, Wang F, Zhu S, Wu Z (2018) Self-Assisted First-Fix Method for A-BDS Receivers with Medium- and Long-term Ephemeris Extension. Hindawi Mathematical Problems in Engineering Article ID 5325034: 1-14.

Shou H. (2014) Orbit Propagation and Determination of Low Earth Orbit Satellites. International Journal of Antennas and Propagation Article ID 903026.

Stacey P, and Ziebart M (2011) Long-Term Extended Ephemeris Prediction for Mobile Devices. ION-GNSS 3235-3244.

Thode HC (2002) Testing for Normality. 1st Ed., Marcel Dekker, New York.

Trujillo-Ortiz A, Hernandez-Walls R, Barba-Rojo K, Cupul-Magana L (2007) HZmvntest: Henze - Zirkler's Multivariate Normality Test. A MATLAB file.
http://www.mathworks.com/matlabcentral/fileexchange/loadFile.do?objectId=17931.

Turner L and Sherlock C (2013) An Introduction to Particle Filtering Lecture Notes.